# Redshift and Energy Conservation

Alasdair Macleod[§]

**ABSTRACT**

It has always been considered a serious error to treat the cosmological redshift as a Doppler velocity effect rather than the result of space expansion. It is demonstrated here that in practical terms this is not the case, and that the apparent distance - redshift relation derived from a Doppler interpretation is reasonably consistent with supernova data (though not as good as standard cosmology with dark energy). The normal Doppler effect is examined in detail and shown to conserve energy as expected. Because of the equivalence between the general relativistic space expansion paradigm and the Doppler effect (as demonstrated) the long-standing problem of energy loss associated with the expansion of the Universe is treated in a similar manner to the normal well-behaved Doppler effect. The mechanism by which energy is conserved with the normal Doppler shift is applied to the cosmological redshift and the energy violation disappears. However, an additional luminosity-dependent recession factor is introduced. The effect on astronomical objects is examined and it is found to add only a small additional redshift to a body generating power by nuclear means but can be very large for objects powered by gravity. A possible connection to the claimed anomalous redshift of quasars is considered.

*V2.0 Changes: Figure 4 altered and conclusions drawn from it modified somewhat; discrepancies with the modelling of Davis and Lineweaver investigated – Figure 5 added.*

## 1. Introduction

It is well established that the cosmological redshift originates from a general-relativistic process with the spectral ratio being a measure of the scale factor of the Universe at the epochs of observation and emission[1]. The apparent motion resulting from the expansion of space cannot be rightly considered as a proper velocity, but there is still a tendency to erroneously apply the relativistic Doppler formula to the measured redshift and derive an equivalent recession velocity, effectively transferring the process into the special relativistic domain[¶]. When the geometrical effect of the space expansion is correctly transformed into an apparent velocity, the value is actually equal to the Doppler velocity when distances are short relative to the dimensions of the Universe because both formulae approximate to the linear Hubble law under that constraint[2]. However it would seem the equivalence can not possibly hold at large distances. In the expanding space paradigm the apparent velocity is permitted to exceed the speed of light (with the point of equality defining the observer horizon). This is not possible in the context of special relativity. It is perhaps because of this obvious and significant divergence that no careful analysis of observational data has yet taken place to quantify the discrepancy at high *z* when the Doppler formula is incorrectly applied in place of the Robertson-Walker metric[¶].

In this paper, the redshift-apparent distance relationship of distant supernovae is examined. Perhaps surprisingly, it is found that the luminosity distance obtained by deriving the Doppler equivalent velocity from *z* and applying special relativity to the propagation process is reasonably consistent with observation (similar to the $\Omega_M=1$, $\Omega_\Lambda=0$ prediction of standard cosmology), with the simplest assumption of a flat and empty Universe and constant expansion; in other words if we are consistently special relativistic. In practical terms therefore, the use of the Doppler velocity expression is an acceptable approximation (and simplification) of Friedman-Robertson-Walker (FRW) cosmology so long as one is careful with the form of the Hubble relation.

---

[§] University of the Highlands and Islands, Lews Castle College, Stornoway, Isle of Lewis, Scotland
email: Alasdair.Macleod@lews.uhi.ac.uk

[¶] Perhaps the reason why this interpretation is so deeply ingrained is the visual appeal of the model of the expanding Universe as a hypersurface with an expansion velocity (and equivalent kinetic energy) working against gravity and large-scale cosmological forces. The model makes it difficult to conceptualise the difference between proper and expansion velocities.

[¶] The paper of Davis and Lineweaver[37] will be considered in the text, but their analysis is believed to be erroneous.

One problematic aspect of the cosmological expansion is the apparent loss of energy associated with the redshift. The effect is particularly bad with cosmological background photons received in the current epoch – they are received with only about 0.1% of their emission energy. Attempts to account for the missing energy within the framework of general relativity have met with severe problems because of the difficulty in defining a local gravitational energy density (gravitational energy cannot be expressed in tensor form). As a result, it is widely accepted that energy is not locally conserved in general relativity[3], although claims are made that energy is globally conserved during expansion. This is in stark contrast to the normal Doppler shift where, as demonstrated in the text, energy is conserved on a photon-by-photon basis. We are surely entitled to demand that photons redshifted by the cosmological expansion similarly conform. Having established a working equivalence between a Doppler velocity and the cosmological expansion, an attempt is made here to apply the Doppler energy conservation 'recipe' to the cosmological shift to explore concepts that may later be transferable to general relativity to recover energy conservation. We avoid the normal unproductive approach of associating the missing energy to unknown field properties; instead, the 'missing' energy is accounted for by an apparent kinematic change at the source to guarantee energy conservation in the observer frame.

Though speculative, the analysis does suggest that energy conservation is possible through the introduction of an additional luminosity-dependent redshift term that evolves according to the mass-to-light ratio of the bound system. It is important to emphasise that this is not strictly speaking an 'intrinsic' redshift because the value varies with observer distance. When applied to cosmological entities, the luminosity redshift is seen to contribute only a small amount to the total cosmological redshift of systems generating energy through nuclear processes but can overwhelm the cosmological effect in systems powered by gravitation such as active galactic nuclei. The actual proportion in this latter case is strongly dependent on the mass of the central black hole, the accretion rate and the radiative efficiency.

**2. The Relativistic Doppler Effect**

The Doppler shift will be analysed in some detail to demonstrate that energy is always conserved, both globally and locally (i.e. for each photon). Comprehensive calculations are included as the results are important to the analysis of the cosmological redshift that is the main focus of the paper.

Consider two masses *A* and B separated by a distance *r* with *A* moving at velocity *v* with respect to *B* (*Figure 1*).

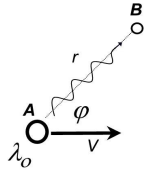

**Figure 1** Relative motion between emitter *A* and absorber *B*

If *A* emits a photon in its rest frame of wavelength $\lambda_o$, the wavelength measured at position *B* is

$$\lambda = \gamma(1 - \beta \cos \varphi)\lambda_o. \qquad (1)$$

$\beta$ is defined as $|v|/c$ and $\gamma = (1-\beta^2)^{-1/2}$. This is the Doppler shift. Visible light may have the wavelength increased towards the red end of the spectrum or decreased towards the blue. The relation is readily derived by Lorentz transforming the system four-vectors and using the invariance of the scalar product of the source and photon four-momenta[5].

The ratio of received to emitted wavelengths in the special case $\varphi = \pi$ defines the redshift factor $z\ (=\Delta\lambda/\lambda_o)$:

$$1 + z = \gamma(1 + \beta), \qquad (2)$$

or equivalently,

$$v = c\sqrt{\frac{(1+z)^2 - 1}{(1+z)^2 + 1}}. \qquad (2a)$$

The relativistic Doppler shift has two components: the Doppler term $(1-\beta\cos\varphi)$ is analogous to the acoustic Doppler effect and is dependent upon angle; the transverse term $\gamma$, the time dilatation factor, has no acoustic analogue and arises because, unlike sound in air, the velocity of light is a constant independent of reference frame. The transverse term is sometimes referred to as the anomalous, perpendicular or second order Doppler effect. The correctness of equation (1) has been established to an extraordinary level of accuracy by a recent experiment[6].

Consider the case of an isotropic light source of luminosity *L* radiating photons of wavelength $\lambda$. Assume a set of stationary observers on the spherical surface at distance *r*. The flux, *f*, measured by each observer is

$$f = \frac{L}{4\pi r^2}. \qquad (3)$$

If the source is instantaneously accelerated to a velocity *v* as shown in *Figure 2*, then each observer notes a change in the flux after a due propagation delay.

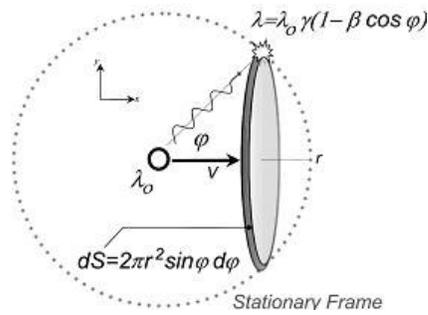

**Figure 2** Light source accelerated to a velocity *v*

Consider first the limited case where $\varphi = \pi$[¶]. The energy flux measured by the observer is reduced by a factor $(1+z)^4$ in comparison with equation (3) (see, for example, Misner, Thorne and Wheeler[1], sections 22.6 and 29.4). The four powers of $(1+z)$ include a combination of local and global effects. The energy of each photon is reduced by $(1+z)$. The received photon count is reduced by the same factor. There is an additional $(1+z)^2$ diminution in flux because the absorber in the moving frame lies on the surface of a sphere of radius equal to the rest-frame separation times $(1+z)$ - the effective photon journey time (distance) is longer by this factor. Of course, this analysis assumes a point source. A real source will appear extended and energy over the full extent of the source must be integrated for the $(1+z)^4$ flux reduction to apply.

The analysis is slightly different for extended sources where the surface brightness is the appropriate parameter rather than the luminosity, for example, the cosmic microwave background radiation (CMBR). In *Appendix A*, it is demonstrated that a blackbody spectrum of a source at temperature $T$ in the rest frame is transformed to a blackbody spectrum of lower temperature $T/(1+z)$ if a source recession velocity as defined by equation (2a) is applied. The fact that the CMBR data is verified to have the form of a blackbody spectrum to high accuracy[7] shows that, for this specific example, the general-relativistic space expansion can be equated to an equivalent proper velocity and treated with special relativity.

Referring to *Figure 2*, the diminution of flux can be generalised for observers at all angles $\varphi$ as $[\gamma(1-\beta\cos\varphi)]^4$. We will determine the luminosity of the moving source by integrating the emitted radiation over the closed surface on which the observers reside. The distortion of the sphere by the Lorentz boost makes the integration a bit messy as the surface is no longer spherical. However, if two of the powers are dropped, the absorption surface is transformed back to a sphere (converting to angular distance). The total luminosity in the moving frame $\acute{L}$ is therefore

$$L' = \int_0^\pi f \frac{2\pi r^2 \sin\varphi}{\gamma^2(1-\beta\cos\varphi)^2} d\varphi. \tag{4}$$

Using the substitution, $u = \cos\varphi$, the integral simplifies;

$$L' = \frac{2\pi r^2 f}{\gamma^2} \int_1^{-1} \frac{-1}{(1-\beta u)^2} du. \tag{5}$$

Integrating and evaluating

$$\begin{aligned} L' &= \frac{2\pi r^2 f}{\gamma^2} \left[ \frac{-1}{\beta(1-\beta u)} \right]_1^{-1} \\ &= \frac{2\pi r^2 f}{\gamma^2} \left( \frac{2}{1-\beta^2} \right) \\ &= 4\pi r^2 f \\ &= L \end{aligned} \tag{6}$$

The last line of simplification follows from equation (3).

Although individual photon energy alters as a result of the velocity boost, the luminosity of the source does not vary. Energy is globally conserved. This does not mean that energy conservation is maintained by some form of collusion between blue-shifted photons with a surplus of energy and red-shifted photons with a deficit: we can show that energy is locally conserved for each photon by a simple kinematic derivation of the Doppler shift expression. For each photon the difference in energy associated with the wavelength shift is actually balanced by a kinetic energy change at the source. Of course, this automatically follows from the transformation of the four-vectors used to derive equation (1) but the alternative derivation below is more illuminating.

---

[¶] This is the angle at which the Doppler effect mimics the cosmological redshift.

Let the source have an equivalent rest mass $M$ and let the photon emission in the rest frame correspond to a loss of mass of $\Delta M$. The equivalent photon energy, $E_{\lambda o}$ is $\Delta M c^2$ and the momentum is $\Delta M c$. From the observer frame the total energy of the source is

$$E^2 = p^2 c^2 + M^2 c^4. \tag{7}$$

Differentiating with respect to $p_\mu$,

$$E \frac{\partial E}{\partial p_\mu} = p_\mu c^2 + M c^4 \frac{\partial M}{\partial p_\mu}. \tag{8}$$

Since the momentum in this case is in the $x$ direction only, only the partial equation in $x$ need be considered. If $E_\lambda$ is the photon energy measured from the moving frame, $\partial E = - E_\lambda$ and because $\partial M = -E_{\lambda o}/c^2$, equation (8) becomes

$$E_\lambda = -v \partial p_x + \frac{E_{\lambda o}}{\gamma}. \tag{9}$$

In the observer frame, the change in momentum in the $x$ direction is $-E_\lambda \cos\varphi/c$. Note that the momentum projection onto the $y$ axis can be ignored because $v_y$ is instantaneously 0 hence has no effect on the source energy (this is why the $y$ and $z$ partials of equation (8) were ignored; see *Appendix B* for further discussion of this point). The equation becomes

$$E_\lambda = \frac{E_{\lambda o}}{\gamma(1-\beta\cos\varphi)} \tag{10}$$

consistent with equation (1) after taking into account that $E \propto \lambda^{-1}$. In effect the photon recoil changes the velocity and kinetic energy of the source, taking up the energy difference. Of course the source must be isotropic if the velocity in the $y$ direction is to remain zero. For an isotropic source, the velocity in the $x$ direction will also remain globally constant from symmetry considerations and the principle of relativity (a Lorentz boost cannot introduce an acceleration). Non-isotropic sources can be created using mirrors but it is shown in *Appendix B* that anisotropic sources also respect energy conservation. The analysis concluded here verifies the Doppler effect conserves energy on a photon-by-photon basis.

### 3. The Cosmological Redshift

The systematic redshift of distant galaxies is interpreted as the expansion of the space between galaxies. If $a(t)$ is the scale factor or size of the Universe at epoch $t$, it can be demonstrated[1] that the cosmological redshift is related to the scale factor by

$$\frac{\lambda}{\lambda_o} = 1 + z = \frac{a(t_{received})}{a(t_{emitted})}. \tag{11}$$

The bolometric flux is given by the following expression

$$f = \frac{L}{4\pi R^2 (1+z)^2} \tag{12}$$

where *R* is radius of curvature of the 2-sphere surrounding the emitter and passing through the receiver at the time of reception. An analysis of the WMAP cosmic background radiation (CMBR) data suggests that the Universe is completely flat[8], hence

$$R = a(t_{received}) \int_{t_{emitted}}^{t_{received}} \frac{dt'}{a(t')}. \tag{13}$$

In principle, the form of the function *a(t)* can be deduced from a comparison of the measured flux and luminosity of standard candles over a range of *z* values. This is objective of the High-Z Supernova Search Team[9] and the Supernova Cosmology Project[10].

Robertson[11] confirmed the $(1+z)^4$ dimming with redshift with the general-relativistic expansion model previously derived by Tolman which is consistent with the present nature of the CMBR[¶].

### 4. Doppler Velocity-Distance Relation for Type Ia Supernovae

It is commonplace to interpret the cosmological redshift as arising from proper motion and to determine the velocity using the Doppler expression of equation (2a). This is clearly incompatible with the general-relativistic description of the cosmological expansion. Harrison[2] in his section 6.1 states that it is not possible to extract the velocity from *z* and relate this to the linear Hubble law $V(z) = H(t)d_e$ (where $d_e$ is the proper distance). However, his reason for rejecting the procedure is based on the incompatibility of the FRW and Doppler velocity models not observational data. In practice, is it really such a bad assumption to make? Are the errors serious? Narlikar[12] and Longair[13] claim the error is fundamental and significant but again fail to quantify the comments. It is difficult to make a simple comparison because the Doppler effect is dependent only on the conditions at the photon emission and absorption points; in the FRW model, changes in the rate of space expansion during propagation affect the redshift hence an integration over history is required to get a parametric solution.

In this section we will take a pragmatic approach to the dispute and try to quantify the error that is actually introduced when the Doppler velocity is assumed by making a comparison with observational data. In this way, we may determine the *z*-domain over which the assumption is acceptable.

We can initially make a rough theoretical calculation and show that for small *z*, the two formalisms are equivalent. Equating the Doppler and general relativistic derivations of *z*:

$$\sqrt{\frac{1+\frac{v}{c}}{1-\frac{v}{c}}} = 1+z = \frac{a(t_e)}{a(t_r)}. \tag{14}$$

The size of the universe at emission time is therefore related to the apparent velocity by

$$v = c\frac{a(t_r)^2 - a(t_e)^2}{a(t_r)^2 + a(t_e)^2}. \tag{15}$$

If we let $a(t_r) - a(t_e) = d_e \ll a(t_r)$ then

$$v \cong \frac{cd_e}{a(t_r)} \cong \frac{d_e}{t_r} \cong H_o d_e, \tag{16}$$

---

[¶] Robertson in the same reference derives the $(1+z)^4$ effect in special relativity as discussed in our Section 2.

which is the Hubble law. We have not been particularly careful with the proper distance or even the current value of the Hubble constant (simply equating it to the inverse of the current age of the Universe) but the approximate equivalence at low *z* is clear.

We are more interested in what is happening at high *z* where a clear divergence is expected and this will be investigated by considering the relationship between the apparent brightness and redshift of type Ia supernovae. The procedure followed is to take the same set of supernovae currently being used to determine the form factor of the Universe in Friedmann-Robertson-Walker cosmology[9, 10] and derive a Doppler velocity from each measured *z* using equation (2a). The photon-travel distance[¶], or retarded distance $d_r$, is found by dividing the apparent or luminosity distance $d_L$ derived from the apparent magnitude by a single power of (1+*z*) to correct for the photon energy reduction and dilution effects (the source is treated special relativistically). We are then interested in the relationship between *v* and $d_r$. At the low velocity limit, it should be reduce to the linear Hubble law.

If the data is corrected for host galaxy extinction and the K-correction is applied, the luminosity distance (in units of Mpc) is obtained using the magnitude-distance formula:

$$m - M = 5\,Log\,[d_L] + 25 = 5\,Log\,[d_r(1+z)] + 25 \qquad (17)$$

*M* is the absolute magnitude and *m* is the apparent magnitude. Riess and his team[9] have provided both new and recalculated *m-M* (distance modulus) values for a large number of supernovae. Knop and the Supernova Cosmology project team[10] give only the apparent magnitude. Data from both groups is used in our analysis, specifically:

**High-Z Supernova Search Team**[9]

(a) A set of 23 low z supernovae with K, galactic, stretching and host-galaxy extinction corrections applied. *Column (e) in Table 5 of the reference.*

(b) A set of 24 recalculated supernovae data from 1995 to 1997 with K, galactic, stretching and host-galaxy extinction corrections applied. *Column (d) in Table 4 of the reference.*

(c) A set of 10 HST supernovae from 1997-2000 with K, galactic, stretching and host-galaxy extinction corrections applied. *Column (d) in Table 3 of the reference.*

**Supernova Cosmology Project**[10]

(d) A set of 46 supernovae where *z* > 0.600 with *m-M* calculated. K and host-galaxy extinction corrections applied. *Table 5 of the reference*. Note that a 'zero calibration' on the distance modulus was not performed.

There is a certain amount of overlap (SN 1995ax, SN 1996cl, SN 1887R, and SN1997ap) that illustrates how the data analysis of each group results in a different final value. The data is plotted in *Figure 3*. A small $M_B$-$M_V$ correction has been applied.

---

[¶] The reason for using the photon travel time is that the normal Doppler effect relies only on the relative velocity at emission and reception. For consistency, an equivalent expansion velocity should rely on the parameters that relate to the photon emission and absorption. In the case of the cosmological expansion, the significant parameter is time. We do not perform a second division by (1+*z*) because the rest separation is not considered relevant.

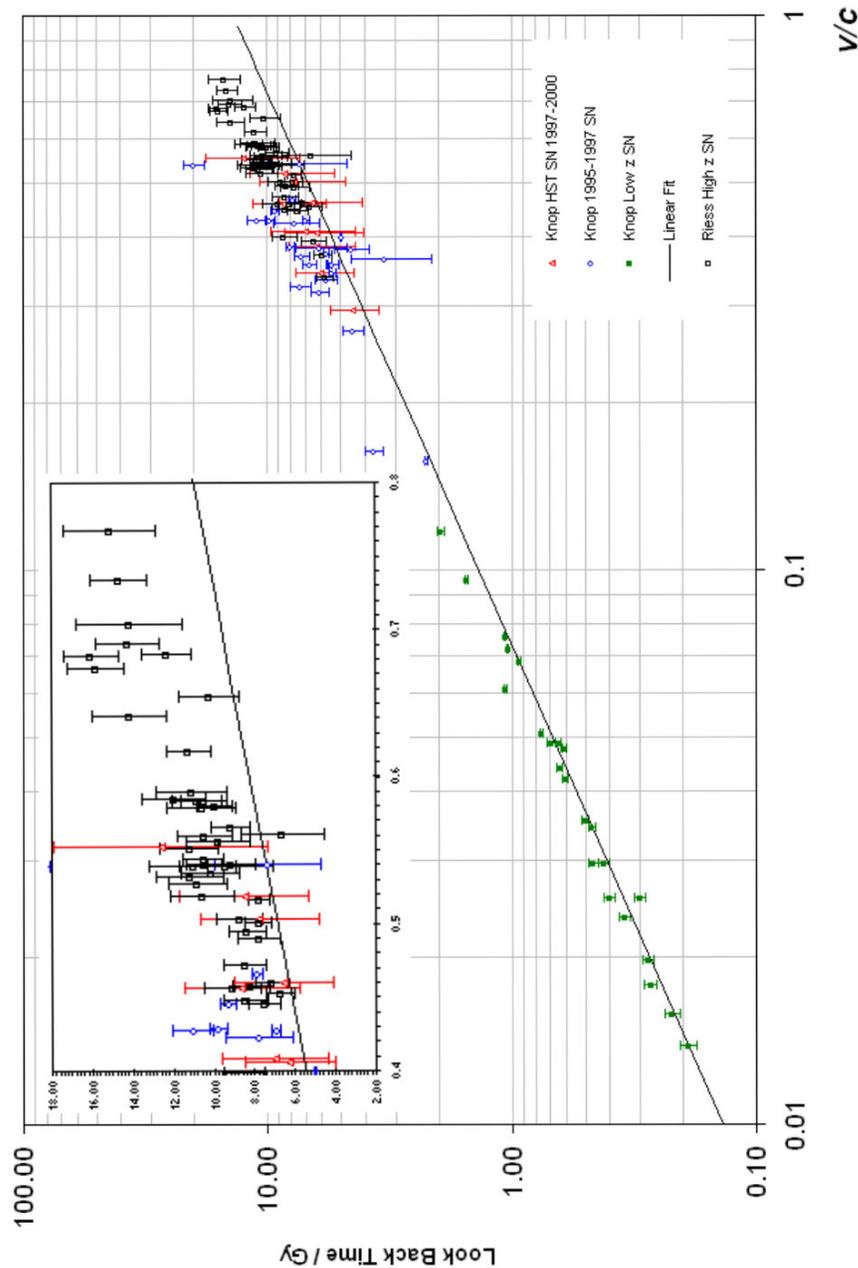

**Figure 3** The look back time (in Gyr) - velocity plot for type Ia supernovae, taken from Knop et al and Riess et al, is displayed with the linear plot expected from special relativity. An absolute magnitude of -19.33 for type Ia supernova is assumed. The solid line represents a universe of age 13.7 Gyr. Inset shows the high-z region with a linear scale.

To model the data, the most straightforward try is to extrapolate the Hubble law relationship and assume the function is linear over the entire domain (although there is no compelling reason why this should be the case). In *Figure 3*, this proposition that the velocity is proportional to the retarded distance, $d_r$, is tested against the dataset. The current value of $H_o$ is taken to be $1/T$ ($T$ is equivalent to the $t_r$ of equation (14)). The Hubble relationship converted to interaction time is therefore

$$v = \frac{c(T - t_e)}{T} \qquad (18)$$

where the age of the Universe is taken to be 13.7 Gyr old (the value that is used for $T$ in the equation). This is shown as the straight line in *Figure 3*. The two points that fall on the extremities of this straight line are easily identified: (0,0) - the velocity is 0 at distance 0; (1, -13.7Gyr) - this is data from the CMBR – the equivalent Doppler velocity is almost $c$ and the retarded time is approximately the age of the Universe.

Equation (18) actually represents a good fit for the data up to $v/c = 0.1$, but supernovae at $v/c = 0.6$ ($z = 1$) appear about 30% further than expected. The prediction is almost identical to that of the Friedmann equation with $\Omega_M=1$, $\Omega_\Lambda=0$. Of course, it was this same failure of the Einstein–de Sitter Universe at high $z$ that required the cosmological term to be reintroduced into standard cosmology.

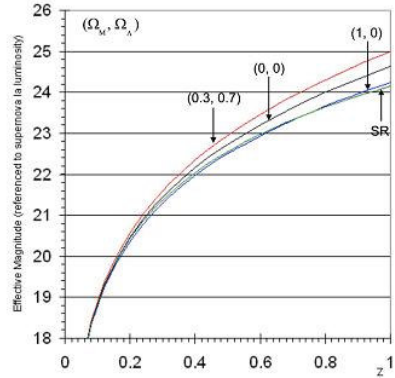

**Figure 4** A comparison of the special relativity (SR) Doppler model and standard cosmology. The results should be compared to Figure 5 of Davis and Lineweaver[37] who place the SR line 23 standard deviations below the $\Lambda$CDM concordance fit (0.3, 0.7) and thus exclude SR as a viable cosmological model.

The peak absolute magnitude is taken to be –19.33 as estimated by Wang[14] (who gives a V-band magnitude of –19.33 ± 0.25). We can conclude that it is legitimate to represent the cosmological expansion with an equivalent Doppler velocity up $z = 1$ with the caveat that there is a significant error beyond $z = 0.1$.

The only adjustable factor in this special relativistic treatment of cosmology is the form of equation (18). It must reduce to the standard Hubble law for $t_e \sim T$, and $v$ should not exceed $c$ as $t_e \to 0$ (only general relativistic cosmology permits the recession velocity to exceed the speed of light). Although we will not investigate further, a modification to the special relativistic Hubble law is justified as equation (18) is not relativistic, and there are many suitable functions that could be explored to enhance the fit (following the precedent of the modification to the basic Friedmann equation to match high redshift data).

*Redshift and Energy Conservation*

Davis and Lineweaver[37] attempt to demonstrate that equations (2a) and (18) do not match observational data. They correctly highlight the difficulty relating the recession velocity to observational features, but plump for the same relationship as our equation (18) here. However, the plot of the special relativity prediction (their Figure 5) is inconsistent with *Figure 3* and appears to be incorrect[§]. *Figure 4* reproduces their calculation with the normalised luminosity distance $H_o d_L/c$ converted to familiar units of magnitude. The special relativity expression (as described above) is $(1+z)[(1+z)^2-1]/[(1+z)^2+1]$. The standard model curves use the expression $z+z^2/2[1-\Omega_M/2+\Omega_\Lambda]$ from Perlmutter[38]. It is clear that the (1,0) case and special relativity are very similar.

Of course, equation (18) has to be largely consistent with all types of observational data, not just distant supernovae. As a further example we can consider the variation of angular width of standard rods with redshift. This is important because proper distance is more difficult to reconcile with the retarded frame used here. For this reason we would expect differences between the Doppler and general-relativistic model predictions.

Consider a 'standard rod' of length $l = 10/h$ pc[¶] (measured with rod and observer at relative rest). There is some doubt as to whether standard rods genuinely exist but some data on radio sources is available[15] and whilst the scatter is large, an averaging process can produce data against which the reasonableness of a cosmology can be tested (although it is unlikely a specific cosmology can be selected).

The angular width in the stationary emission frame is

$$\phi = \frac{l}{d_e}, \qquad (19)$$

where $d_e$ is proper separation distance. Following a velocity boost (in the Doppler formalism), the observer maps the stationary circumference $2\pi d_e$ to the circumference $2\pi d_e(1+z)$ onto which objects are projected hence the apparent length is scaled by a factor $(1+z)$. The distance is also scaled by $(1+z)$ so the apparent angular width remains unchanged. The distance $d_e$ can be derived from observables using a modified equation (18): $d_e = vT/(1+z)$. Converting the angular measurement units from radians to *mas*, and taking $l$ to be 14.0 (= 10/0.714) pc,

$$\phi = \frac{0.671(1+z)}{v/c}. \qquad (20)$$

This function is plotted in *Figure 5* below along with the binned data from 330 compact radio sources gathered by Gurvits, Kellermann and Frey[15].

---

[§] It is probable the $(1+z)$ term in the special relativity model to convert from effective to luminosity distance has been neglected.

[¶] Note that $h$ is derived by expressing the Hubble constant as 100 $h$ km s$^{-1}$ Mpc$^{-1}$, the value being 0.714 for a Universe of age 13.7 billion years

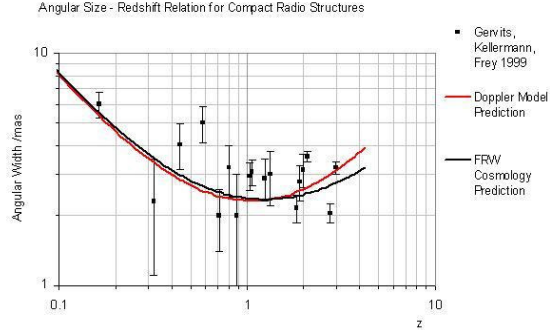

**Figure 5** The red line shows the apparent angular width of a standard rod of length 14 pc as a function of redshift. The match with the data is similar to that of the FRW model.

The fit is reasonable. Equation (20) can be rewritten

$$\phi = \phi_o \frac{(1+z)((1+z)^2 + 1)}{((1+z)^2 - 1)}. \tag{21}$$

There exists a minimum ($d\phi/dz=0$) at $z = 1.06$. In contrast, the FRW cosmology functions are typically of the form

$$\phi \approx \frac{(1+z)^2}{(1+z) - \sqrt{1+z}} \tag{22}$$

with a minimum at $z = 1.25$. This is shown for comparison in *Figure 4*. Of course, for the poor quality of data available, the expressions cannot be distinguished and the equivalence between the Doppler and general relativistic formalisms still holds.

### 5. The Cosmological Redshift and Energy Conservation

The cosmological expansion appears not to conserve energy. Photons are received with a lower energy than when emitted and because the source appears to be moving away from all the distant absorbers, there is no obvious mechanism by which energy can be conserved – all photons lose energy. Ideas involving a variation in gravitational energy or zero-point energy as space expands tend to fail because the expansion is determined by matter density and to balance the energy loss we would have to postulate local variations in expansion based on photon density. Whilst these ideas are very appealing and evolve fascinating space-time topologies, they ultimately fail to satisfy energy conservation requirements.

Exploiting the working equivalence between a Doppler velocity and the expansion and using equation (18), a previous paper[16] argued that energy could be conserved from the viewpoint of the emitter if a distant absorber subject to the cosmological expansion were attributed an 'Hubble Energy'. It was shown that this could give rise to anomalous accelerations similar to that experienced by the Pioneer probes. In this paper, we will instead consider the situation from the viewpoint of the absorber and explore the possibility that the supposed energy loss associated with the cosmological redshift can be dealt with in a manner analogous to the Doppler shift described earlier.

Consider an object subject to the cosmological expansion. The radiation power loss is proportional to the luminosity *L* and the observer distance. If we consider the source 'surrounded' by a set of observers on the surface of a sphere of radius *r*, the total integrated power loss is $zL/(1+z)$. The apparent momentum gain at the source is $L/c(1+z)$ away from the observer. The assumption is made that the momentum associated with all the photons can be added in a scalar fashion, even though they are emitted in different directions – this means that even if the output is beamed, the cosmological momentum change is the same for all observers regardless of the angle of orientation with respect to the beam. By analogy with the Doppler equation, we can postulate an increase in the recession velocity equivalent to the momentum change. Consideration of equation (8) shows this procedure conserves both energy and momentum. The increase in cosmological velocity, per second, is

$$\Delta v = \frac{L}{Mc\gamma(1+z)} = \frac{L}{Mc}(1-\frac{v}{c}). \tag{23}$$

All velocities are assumed to be cosmological. We can get an idea of the accumulated effect by assuming the source has had the same mass-to-light ratio from $T = 0$ up to the time the photon was emitted, $T_e$. Multiplying $\Delta v$ by $T_e$ and using equation (18), the accumulated increase in velocity is

$$v_l = \frac{280}{\Pi}(1-\frac{v}{c})^2 \quad km\ s^{-1}. \tag{24}$$

The subscript $_l$ indicates an 'luminosity' velocity which gives rise to a 'luminosity redshift'. The value of $T_o$ is taken as 13.7 Gyr and $\Pi$ is the mass-to-light ratio in units of $M_\odot/L_\odot$.

This type of crude calculation would be appropriate to relatively near galaxies. The mass-to-light ratio by galaxy type is remarkable constant. Faber and Gallagher[17] give estimates for a range of galaxy morphologies. For example, the value of $\Pi$ for a type $S_{cd}$ galaxy is 3.9. Slotting this into equation (24) and assuming *v/c* is small, the increase in velocity is about 72 km/s. S0⁻ galaxies by comparison give a value of about 28 km/s. This might offer an explanation for the tendency of smaller galaxies (Sc I) to have a systematically larger redshift[18]. Note that the effect cannot be related to the observation that young, blue stars to have an anomalous redshift[19]– stars in our galaxy do not have a cosmological velocity component therefore there can be no luminosity velocity component. It should be noted that luminosity velocity component can decay with time as the galaxy attracts non-luminous matter such as hydrogen gas or dark matter – the velocity drops through conservation of (cosmological) momentum: $m_1v_1 = [m_1 + \delta m][v_1 - \delta v]$ where $\delta m$ and $\delta v$ are both positive.

To summarise, energy and momentum conservation require that distant objects have an additional cosmological velocity component proportional to their integrated luminosity. The effect on galaxies is small but measurable and is generally distance independent. The effect would show up as a discrepancy between the redshift and Tully-Fisher or supernova calculated distances.

The additional velocity cannot be associated with to a physical change in position (radial distance) otherwise the ordering of events can be changed in an acausal way. It therefore follows that the source is at the position indicated by the cosmological element of the recession velocity and that this additional luminosity component be intrinsic to the observer (not the source). In what sense can it therefore be considered a velocity? This question is left open for the moment. Remember the luminosity component represents energy conservation and the actual mechanism is unimportant – any mechanism with these conservation properties will have the same observable effect.

**6. Cosmological Redshift and Quasars (QSOs)**

For some systems it more appropriate to consider the total energy loss over the lifetime of the source rather than on the basis of small time intervals. Let $M$ be the original total mass of the source, and $\xi$ the fraction of mass eventually converted into energy. The total increase in intrinsic velocity is of the order

$$\frac{v_l}{c} = \frac{\xi}{1-\xi}(1-\frac{v}{c}). \tag{25}$$

For a compact or extended source (i.e. a star or galaxy) powered by nuclear reactions, $\xi$ has a value of about 0.001 and the redshift change is of the same order (consistent with the normal galaxy mass-to-light ratio applied to equation (24)). The situation is very different in the case of an accreting black hole (BH). The power generation mechanism is then gravitational and $\xi$ can then be identified as the radiative efficiency, a factor whose value is typically believed to be about 0.1 but can approach 1, the latter case being where all incident mass is converted to energy. As an example, if we consider a BH accreting with a radiative efficiency of 0.9, and with a cosmological value of $z$ of 0.1 (based on distance), the measured redshift would be $z \cong 3$. It is therefore possible that active galactic nuclei may have a large redshift associated with luminosity mixed in with the cosmological redshift[¶].

The most visible evidence of galactic core activity is QSOs. There has been a long running dispute over the nature of QSOs and whether they are really located at the redshift distance. The main points of each side of the argument will be considered and we will judge whether the notion of a luminosity redshift is consistent with observational data and can help resolve the dispute with the standard cosmological model[§].

The first point that strongly favours the standard interpretation of AGNs in general and QSOs in particular is that there exists a reasonable theory that not only explains many of the observations but actually unifies apparent disparate objects into a single schema[20]. However, the observational data available is limited, and the model has not been rigorously tested. It is rather difficult to make redshift comparison over the full extent of an active galaxy: the jet frequently associated with AGNs is often the most visible component but generally produces radiant energy by synchrotron radiation - there is no spectral information from which to obtain a redshift. Certainly the redshift can be extracted from the core emission, the source being the central accretion zone, but only occasionally can light from the host galaxy be resolved. This fact that the host galaxy can sometimes be resolved is again strongly suggestive of QSOs being truly distant objects. On the occasions when resolution is possible, there is no example of the host galaxy having a different redshift to the central active region. This indicates there is little gravitational redshift associated with emission from around the central black hole and severely restricts alternative explanations. A QSO lifetime of $10^6 - 10^8$ years is estimated[21], activity presumably beginning as gas coalesces to form the galaxy and finishing when the nearby matter has been consumed. The supply of additional material through galactic collisions and mergers is thought to reactivate the process later in the lifetime of a galaxy.

We can immediately discount alternative models that consider all QSO redshift to be intrinsic with the sources actually stars in the galaxy - the idea presumably arises from the similarity to jets from young stars and micro-quasars in the galaxy. The idea fails because the source distribution is generally isotropic with no correlation with the galactic plane. Arp has suggested an alternative mechanism for QSO activity based on observed QSO-galaxy associations. In Arp's model[22], QSOs are ejected from parent galaxies, initially with a high intrinsic redshift, but with the redshift declining with time as the quasar evolves into a normal galaxy. The idea is given some theoretical credence by the notion of particles being created at galactic cores with initially zero mass but acquiring their normal mass over time. Part, but not all the redshift is intrinsic by this model.

---

[¶] Note that it is not just the active nucleus that has the additional redshift. It affects all the bound mass, including the host galaxy. The conservation redshift cannot be determined by comparing the host and core redshifts.

[§] Care is taken to avoid the definition of an intrinsic redshift as the value is relative to the observer and is thus not strictly intrinsic.

Comparing the standard and Arp's idea, an HST survey of 20 near-QSOs[23] has resolved the host galaxies. Most galaxies are normal and do not show evidence of the mergers that would be expected to trigger late QSO activity. However many had galactic companions close by (at about the same redshift), which is largely consistent with Arp's model, but the authors interpreted the associations as a merging rather than diverging process. Luminous QSOs are associated with a young host stellar population[24] and there is a tendency for the host to be brighter than normal galaxies, again in accordance with Arp's observations. Recent attempts[25] failed to resolve host galaxies of quasars at $z \sim 2$. This would be expected from Arp's model where these are considered very young objects that have not yet formed galaxies.

One problem with Arp's model is that if this is the way galaxies are produced then all QSOs should be associated with galaxies. In fact the anomalous associations affect only a very small proportion of objects. Is it reasonable to expect there to be two very different mechanisms for quasar production to be at work? Another problem is that the supporting theory of mass creation is far removed from established physics with no rationale for the process in the standard particle model or indications of it from high-energy experiments. As such there is no corroborating evidence.

However Arp's observations do appear to be reasonable and should not be dogmatically rejected, and in fact the limited anomalous data is readily incorporated into a standard model augmented with luminosity redshift. We may consider one process whereby galaxies may form to show how the anomalies might arise. Active nuclei eject huge volumes of material through the jet and counter jet often present. As the ejecta is slowed by the interstellar medium, it is natural to assume that gravity will draw the matter together. We would expect a BH to rapidly form and efficiently accrete, appearing as a QSO. Peripheral gaseous material is gravitationally attracted but not initially gravitationally bound; the mass-to-light ratio of the luminous source will be core dominated and therefore very low, resulting in an anomalously large redshift (by equation (25)). As peripheral material loses kinetic energy through collisions, the bound mass will increase (without a corresponding increase in energy output) hence the mass-to-light ratio will fall. We might therefore expect galaxies to begin as QSOs with an initially high anomalous redshift that decays to that of a normal older galaxy over time as the body of the galaxy forms and stars begin to shine - perfectly consistent with Arp's observations. The short duration of QSO activity would indicate that it contributes little to the overall luminosity redshift over the lifetime of the galaxy and we would expect to see little residual redshift in older nearby galaxies. A second phase of QSO activity would not result in a significant anomalous redshift. By this process, we might expect to observe a small number of quasars caught in the early act of galaxy creation with a significant anomalous redshift but the vast majority in the later stage of quasar activity or a later triggering with little or no anomalous redshift. This is consistent with observation and for the most part QSOs will be at the redshift distance.

This is very speculative of course, but the notion of galaxies ejecting material, the ejection material forming galaxies which in turn eject, and so on, may help explain the large-scale structure of clusters and super-clusters. In this scenario the collective streaming motion of galaxies is the residual velocity of the jet from which the cluster formed. The QSO peak at around $z=2$ and the varying intrinsic luminosity can presumable be modelled in a manner analogous to standard growth-supply equations.

Other evidence has been presented for QSOs being at the redshift distance. One of the best examples is a study by Qin, Xie, Zheng and Liang to compare the emission and absorption lines of 400 quasars[26]. They found that all quasars had one or more absorption lines with a lesser redshift. This is interpreted as absorption by intervening galaxies or gas clouds and is strong evidence for QSOs being at the redshift distance. The analysis does not and cannot disprove the possibility of a fractional non-cosmological redshift in a small number of cases. As with all QSO surveys, the analysis turned up a surprise – 16% of the QSOs had absorption lines at a greater redshift then the emission line. The maximum $z$ difference was 0.1450 and the mean was 0.0137. One of the few ways to account for this is to postulate that QSOs have a high proper velocity relative to the local neighbourhood, presumably as a result of the material ejection. The QSOs displaying this effect will be those moving towards the observer, the percentage being amplified by the selection effect.

Following this line of reasoning, it should not be surprising to find QSOs associated with superluminal motion and proper motion. In the standard model, QSOs are not supposed to exhibit proper motion (and indeed this assumption forms the basis of an effective search technique), but if they are indeed associated with galactic ejection, limited proper motion might be expected in many QSOs, even accepting the huge distance. An analysis of 580 radio sources from 1980 to 2002[27] concluded that most exhibited proper motion, one as large as 0.4 mas/yr but typically about 0.05 mas/yr[¶]. At $z=2$, where the redshift is cosmological, 0.05 mas/yr corresponds to 2.6 light years per year. This is more than expected, but the paper does not make it clear how the QSO core is distinguished from ejected material (which is known to emit at radio frequencies and move superluminally) from which the radio energy may really originate.

QSO output is known to vary significantly over periods as short as hours. If there is no evolution of variability, one would expect the time scale to show a $(1+z)$ dependence. This is true even if part of the redshift is a result of the conservation mechanism described - the special relativistic dilation and $(1+z)^4$ dimming are derived from the nett velocity. The finding of Hawkins[28] that there appears to be no $(1+z)$ dependence is at odds with the standard model and the conservation modification. A larger survey of 25,000 quasars from the Sloan Digital Sky Survey[29] also failed to detect a $(1+z)$ dependence.

Gravitational lensing is a strong indicator that QSOs are at the redshift distance (e.g. QSO 0959+561). In many cases, the intervening lensing galaxy or matter concentration has been detected.

One other intriguing aspect of QSOs is the claim that the redshift is quantized. There are two schools of thought; the first believe that the redshift difference between galaxies and nearby, presumably ejected, QSOs is a multiple of 0.06. This seems to be discredited by an independent survey that searched for a $Log(1+z_{difference})$ periodicity in 1647 QSO-galaxy pairs identified in the 2dF quasar survey[30] and found none. Bell proposes a much more complex scheme involving sums of series (but still based on the $z$ value 0.062) and is supported by reasonable evidence[31, 32].

If is incredible to think that quantization could be possible. The observed redshift may have a continuous cosmological component, a proper motion component (continuous because of the angle of orientation with respect to the observer) and possibly even a gravitational component[33]. Added to the mixture is the possibility of a luminosity redshift component. How can quantization possibly occur? Rather than reject the possibility out of hand, we can postulate that quantization is possible only if one of the elements is quantized and dominates all other factors. The cosmological and proper motion components cannot be quantized. It is difficult to accept that the gravitational redshift is significant because the core and host galaxy would have different redshifts so the last possibility is the luminosity redshift. Its value depends on BH dynamics, growth and accretion rates and possible preferred rates may occur, but the problem is that the luminosity redshift is not intrinsic but relative to the observer. For this reason, we must be dubious that it can offer an explanation for the claimed quantized redshift observations.

A redshift – apparent magnitude plot of 64,866 QSOs, BLac objects and Active Galaxies taken from an online catalogue[34] is shown in *Figure 6*. There is no obvious indication of quantization, but it does highlight the problem many of the dissenters have with the standard model of QSOs. From $z = 0.4$ to $z = 3$, the apparent magnitude and the dispersion is uniform. There is no hint of the diminishing brightness associated with Hubble's law and obeyed reasonably well by galaxies and supernovae. Note that the diagram includes a $(1+z)^4$ flux reduction hence the absolute magnitude increases dramatically with redshift. Presumably, the critics question the coincidence of the $z$-independent apparent magnitude but standard theory explains this by evolution and it is consistent with the notion of galaxies formed from ejecta – the volume of material is divided with each step of the cycle and hence the luminosity will decline. As such, we might find any luminosity redshift mixed in with many QSOs at the actual redshift distance with huge luminosity and simply add to the distribution but the factor is probably too small to significantly skew the distribution.

---

[¶] The observations in the reference are in the range $z = 0 – 4.4$. These observations will be subject to the angular width relation of Figure 4 and with the assumption of a consistent proper motion could form another test of the cosmological models.

Although there are many questions, the standard model is the best description of active nuclei currently available and is generally adequate. There are significant anomalies, and while they represent only a fraction of the complete evidence, an explanation is still required for each and in the most part this are not forthcoming. The introduction of a luminosity redshift helps to explain some of the anomalous data but by no means all. An analysis of the detailed properties of BHs and accretion during galaxy formation is required to make detailed quantitative predictions with which the idea can be properly tested. In conclusion, we will cite the case of galaxy NGC 7603 and its surroundings to show that the dissenters do have some fair ammunition against the standard model and the objections will not simply go away[¶].

Galaxy NGC 7603 ($z = 0.029$) was investigated by Sharp in 1986[35]. An extended spiral arm ends on a smaller, higher surface brightness galaxy NGC 7603B with a redshift of 0.057. It was commented that the simplest explanation is of interaction, but because of the disparate redshifts this is clearly impossible. Sharp concluded that it is a chance projection effect. A more detailed investigation by López-Corredoira and Gutiérrez in 2004[36] showed two faint galaxies with $z = 0.245$ and $z = 0.394$, one at each end of the narrow connecting filament. The filament had a redshift consistent with that of the main galaxy over its entire length. The arrangement of the group is striking because of the agreement with the notion of galaxies forming at knots in jets, initially with a high redshift because of energy conservation.

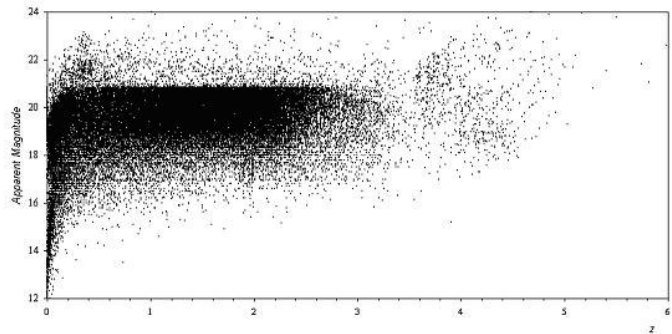

**Figure 6** A plot of the apparent magnitude against redshift for all the objects in the Véron-Cetty & Véron catalogue of QSOs and active galaxies (11th Edition, 2003)[34]. All the 2dF and the first part of the SLOAN survey results are included. The plateau arises from the 21 magnitude sensitivity limitation of some of the surveys, and the weak striation is a rounding effect.

---

[¶] There are many examples give by Arp, E M Burbridge, G R Burbridge, Hickson and many others of apparently anomalous redshift associations between galasies or QSOs and galaxies. There is a considerable literature on the subject.

**7. Conclusion**

In this paper the issue of energy conservation associated with the normal Doppler shift of light was examined in some detail and was found to hold. The cosmological redshift was examined and it was shown that it could be treated as a Doppler effect arising from an equivalent velocity. Having established a working equivalence, an attempt was made to re-establish energy conservation with the cosmological expansion by treating it like a Doppler effect with an additional recoil 'cosmological velocity'. The new redshift may explain some reported anomalies in the redshift of QSOs.

**Appendix A**

**The Cosmic Microwave Background Radiation (CMBR)**

The CMBR can be given a Doppler interpretation in a way that is consistent with the observation that the transformed blackbody source remains a blackbody when viewed today. Referring to *Figure 2*, the cosmic background clearly cannot be treated as a point source as it surrounds the observer. The picture has to be reversed. Previously the source was at the centre and the observers were on the surface of the sphere. In this case, the observer is in the centre and the emitters are on the surface. When there is no relative velocity, the emitters appear on a continuous surface of a sphere of radius *r*. The moving observer is centred in a circle of radius $(1+z)r$, assuming a special relativistic treatment of the recession velocity. The radiance or surface flux is the total energy emitted in unit time from unit surface area at the emission 'surface' in the normal direction (units: $Js^{-1}m^{-2}$). For a black body, the radiance is found to be related to the surface temperature by Stefan's Law:

$$R = \sigma T^4. \qquad (A1)$$

This is the actually the integral of the blackbody energy density spectrum:

$$\rho_T(\lambda)d\lambda = \frac{8\pi hc}{\lambda^5} \cdot \frac{d\lambda}{e^{\frac{hc}{\lambda kT}} - 1}. \qquad (A2)$$

σ is Stefan's constant and the integration of equation (A2) gives

$$\sigma = \frac{2\pi^5 k^4}{15 c^2 h^3}, \qquad (A3)$$

as has been experimentally established.

If the observer is assumed stationary with respect to the CMBR surface of last scattering, the measured radiance at a detector is the same as the surface radiance because the observer is bathed in the equilibrium photon gas. If the observer is moving at velocity *v* with respect to the scattering surface we can show that the blackbody spectrum of temperature *T* is transformed to another blackbody spectrum with a lower temperature $T/(1+z)$.

The measured radiance will be reduced by $(1+z)^4$ because of the reduction in photon energy and photon number and the reduced energy density at the emitting surface (the last scattering surface is projected onto a greater surface area hence the surface density of emitters is reduced, not exactly the same argument as for the point source). The measured energy density between wavelengths λ and λ+dλ is

$$\rho_\lambda^T d\lambda = \frac{\rho_{\lambda/(1+z)}^{(1+z)T}}{(1+z)^4} \frac{d\lambda}{(1+z)} = \frac{1}{(1+z)^5} \frac{8\pi hc(1+z)^5}{\lambda^5} \frac{d\lambda}{(e^{\frac{hc(1+z)}{\lambda kT(1+z)}} - 1)} = \frac{8\pi hc}{\lambda^5} \frac{d\lambda}{(e^{\frac{hc}{\lambda kT}} - 1)}.$$

$$(A4)$$

This calculation shows that the observed energy is that of a black body of temperature *T*. The energy is consistent with having originated from a blackbody of higher temperature $(1+z)T$ receding from the observer at the velocity extracted from *z* as in equation (2a). The result is consistent at all wavelengths [¶].

**Appendix B**

**Anisotropic Doppler Source**

It is worth examining if it is possible to violate energy conservation by creating a source that is not isotropic. If, for example, a mirror is placed behind the source and accelerated with it (*Figure B1*): Is energy still conserved?

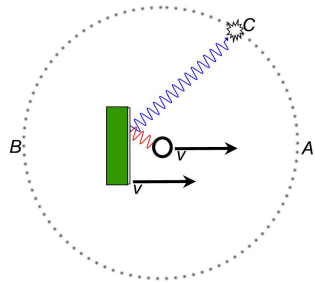

**Figure B1** Light reflected from moving mirror

Note first that if the mirror is located at position B, stationary, then to observer A, looking at the source through the mirror will see the source moving away at velocity *v*, and thus with the same energy that an observer at B would measure. Note that the photons will exert a force on the mirror through their momentum, but we will assume the mass is so great that no energy is transferred. If this mirror at position B (or any place between B and the source) is instantaneously accelerated to velocity *v*, an observer at, for example, position C will see the source moving towards then through the mirror. The photons that were emitted with a red shift should be measured to have a blue shift. The concept of the mirror 'doing work' on the photons is fraught with conceptual difficulties hence we will simply conduct a kinematic analysis (energy and momentum conservation). The analysis differs from the analysis of the source because the rest mass of the mirror is not altered – it can gain or lose momentum and kinetic energy – that's it. Differentiating the energy equation for the mirror (equation (7)), we obtain

$$\frac{dE_{mirror}}{dp_{mirror}} = v_{mirror} = v \tag{B1}$$

Consider that the incident photon of energy $E_{rest}/\gamma(1-\beta cos\varphi)$ which hits the mirror and emerges with energy *E'*, imparting the change of momentum *dp* on the mirror in the process. In the *x* direction, we can work out the final energy and the change in momentum

$$E' = \frac{E_{rest}}{\gamma(1-\beta\cos\varphi)} - dE_{mirror}, \tag{B2}$$

---

[¶] The transformation is consistent with Stefans law and also with Wien's Law which states that $\lambda_{max} * T$ is a constant.

$$dp_{photon} = \frac{\cos\varphi \, E_{rest}}{\gamma(1-\beta\cos\varphi)c} + \frac{E'\cos\varphi}{c}. \tag{B3}$$

Rearranging equation (B2) and dividing by the negative of equation (B3),

$$\frac{dE_{mirror}}{dp_{mirror}} = \frac{E' - \dfrac{E_{rest}}{\gamma(1-\beta\cos\varphi)}}{\dfrac{\cos\varphi}{c}\left(E' + \dfrac{E_{rest}}{\gamma(1-\beta\cos\varphi)}\right)} = v. \tag{B4}$$

This simplifies to

$$E' = \frac{E_{rest}}{\gamma(1+\beta\cos\varphi)}. \tag{B5}$$

This is consistent. The mirror is slowed down supplying KE to increase the photon frequency. The emitter appears to be moving towards the observer at C when viewed through the mirror. With $\varphi = \pi$, the change of a red shift to a blue shift is consistent with this viewpoint. Specifically, the emitter appears to be moving away from the observer when viewed through the mirror. Note that the mass of the mirror is not relevant to the calculation. There is a tiny momentum transfer in the *y* direction because the incident and reflected photon energies are not the same, but the effect cancels over the whole apparatus. Looking again at *Figure B1*, if the mirror is part of the moving assembly, it is clear the now directed output beam will slow down the assembly ensuring energy conservation. There is approximately –2p of momentum associated with the mirror, +p associated with the escaping photon and +p from the recoil of the original emission which has been neglected from the calculation by presuming $E_{rest}$ incorporates any energy loss associated with the emission recoil. Note also the important kinematic difference between absorption and reflection. You may be concerned that the photon wavelength changes without being absorbed by the mirror but we can consider it to be associated with the path length and how it changes.

This is all relevant to the idea of the 'solar sail', a propulsion concept to extract energy from solar photons. There has been some dispute as to whether the concept is feasible but it clearly is: the mirror- emitter apparatus must recoil with a balancing momentum to that of the direct photons. Probably the source of confusion is when we consider the situation from rest frame of the source. The mirror is initially stationary with respect to the emitter therefore the photon reflected back must have the same frequency as that emitted (emitter looks stationary in the mirror hence no redshift). There is no energy transfer so how can the mirror ever begin to move? The energy transfer from a single 'standard' photon of energy $E_\gamma$ as a function of mirror velocity is

$$\partial E = \frac{2v}{v+c} E_\gamma. \tag{B6}$$

Initially the mirror velocity is 0 hence no energy is transferred. The problem is related to Zeno's paradox of motion and the resolution is to recognise that motion is not a continuous but a discrete process. Substituting equation (B6) into equation (8) and noting that the change in internal energy is 0 for a reflection process we obtain

$$\partial p = \frac{2}{v+c} E_\gamma. \tag{B7}$$

The efficiency of the mirrored solar sail doubles as the velocity approaches *c*. For the alternative of a solar sail made up of perfectly absorbing material the equivalent expression is

$$\partial p = \frac{1}{c} E_\gamma. \tag{B8}$$

The change in internal energy is not 0 in this case hence an absorbing sail will heat up and radiate isotropically but with no effect on the nett velocity. If a reflector is placed in the forward direction of the absorbent surface, some of the radiation can be used to increase trust but the efficiency will only approach, not exceed, the performance of a perfect mirror as defined by equation (B7). Note that as the velocity approaches *c*, an absorber and a mirror become equally efficient.

As a final thought on mirrors, consider an observer sending out photons to an orthogonal mirror moving radially at velocity *v* (the positive direction being towards the observer). The kinematic calculation following the lines of the ones above shows that the emitted and received wavelength are related by

$$\lambda_{reflected} = \frac{(1+\beta)}{(1-\beta)} \lambda_{emitted}. \tag{B9}$$

Effectively the observer appears to himself to be moving at a velocity *v* plus *v* *added relativistically* : $2v/(1+v^2/c^2)$. Substituting this velocity into equation (1), with $\varphi = 0$ gives equation (B9). Note that if the mirror is moving transversely, *v* is 0 and from equation (B1) the energy change at the mirror must be 0 – there is no redshift; because the photon path is not altered.